\documentclass[twoside,10pt]{article}
\usepackage{1}
\title{\Large \bf LATEST RESULTS FROM ALICE}
\author{Eugenio Scapparone on behalf of the ALICE Collaboration}
\date{}
\begin{document}
\maketitle

\begin{center}
\vspace*{-0.3cm}
{\it  Istituto Nazionale di Fisica Nucleare,Sezione di Bologna(Italy)\\
}
\end{center}
\vspace{0.3cm}
\begin{center}
{\bf Abstract}\\
\medskip
\parbox[t]{10cm}{\footnotesize
In this paper selected results obtained by the ALICE experiment at the LHC will be 
presented. Data collected during the {\rm pp} runs taken at $\sqrt{s}$=0.9, 2.76 and 7 TeV 
and Pb-Pb runs at $\sqrt{s_{\rm NN}}$=2.76 TeV allowed 
interesting 
studies on the properties of the hadronic and nuclear matter: proton runs gave us the possibility 
to explore the ordinary matter at very high energy and up to very low  
$p_{t}$, while Pb-Pb runs provided spectacular events where several thousands of particles produced  
in the interaction
revealed how a very dense medium behaves, providing a deeper picture on the quark gluon plasma(QGP)
chemical composition and dynamics.}
\end{center}

\section{Detector description} \label{s1}
The most important
requirements for a general purpose heavy ion experiment at the LHC
are a powerful particle identification over 
a wide momentum interval, a robust tracking capability in a very high multiplicity 
environment and a very low cut in transverse momentum $p_{\rm t}$. 
The ALICE detector matches these needs using several detectors, and implementing almost all the known 
particle identification(PID) techniques.  
A robust tracking and the vertex finding are provided  
by the Internal Tracking System (ITS), made by three different silicon based detectors,
followed by a large volume TPC\cite{alicedet}.
As fas as the PID is concerned,
each detector covers a different range of momentum: 
d$E$/d$x$ vs $p_{\rm t}$ is provided by ITS and by TPC, particle masses difference, reflecting 
in a different time of flight and Cherenkov angle, is measured by the Time Of Flight(TOF) and by 
the High Momentum Particle Identification System (HMPID). 
Electrons are tagged by the Transition Radiation Detector (TRD), $\gamma$-rays by the 
Photon Spectrometer(PHOS), 
electromagnetic shower by the EMCAL, 
while muons from the forward muon arm.
The barrel, immersed in a 0.5 T solenoidal magnetic field, is made of several detectors: 
the Silicon Pixel Detector(SPD), the closest to the interaction point (IP), is 
followed by the Silicon Drift Detector (SDD) and by the Silicon Strip Detector (SSD). 
These dominate the vertex resolution, ranging from 250 $\mu$m ($p_{\rm t}$=0.2 GeV/$c$) 
to 20 $\mu$m ($p_{\rm t}$=10 GeV/$c$).
\begin{figure}[h!]
\begin{center}
\includegraphics[width=8cm]{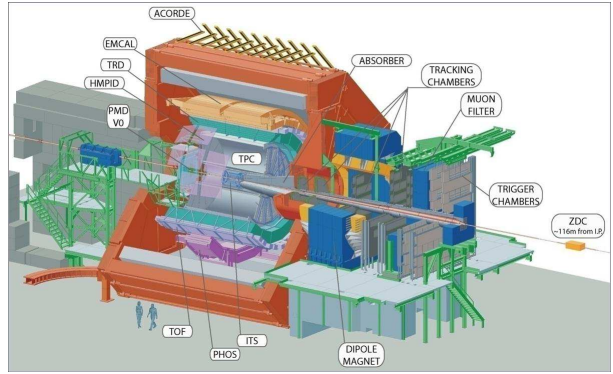}
\end{center}
\vspace{-0.5cm} \caption{Sketch of the ALICE experiment
 } \vspace{-0.5cm}
\label{fig:1}
\vskip 0.5 cm
\end{figure}
Placed at a radius ranging from 0.85 to 2.5 m,
the large TPC, consists of two 2.5 m long drift volume  ($\simeq$88$m^{3}$), separated by a 
central cathode. The large number of samples allows the TPC to measure the track d$E$/d$x$ with a 
 5$\%$ error. 
The TOF is a high segmentation MRPC detector ($\simeq$ 150,000 channels) 
with an excellent time resolution, better than 100 ps, placed over the full
azimuthal angle and $|\eta|$<0.9. It provides a 3$\sigma$ discrimination for $\pi$/{\rm K} 
and {\rm K/p} 
up to 2.5 GeV/$c$ and 4 GeV/$c$ respectively.
The goal of the TRD detector, made of a radiator and a drift chamber operated 
with {$\rm Xe/CO_{2}$} mixture (85$\%$/15$\%$), is the tagging of the electrons. 
The PHOS, the EMCAL and the HMPID are
additional detectors with partial coverage of the central barrel. 
The largest detector in the forward region is the ``muon arm'', consisting
of an hadron absorber, MWPC chambers, a 0.7 T dipole followed by RPCs to tag muons and to and measure 
their $p_{\rm t}$. Several small and important detectors run in 
the forward and backward region and close to the beam pipe: 
T0, V0, Photon Multiplicity Detector (PMD) and the Forward Multiplicity Detector(FMD). Spectator protons and neutrons are detected by
the Zero Degree Calorimeters (ZDC), consisting of two sets of calorimeters placed at 
$\pm$ $\simeq$116 m from the IP.   
More information on detector design and performance can be found in \cite{alicedet}.
\section {Results from Proton-Proton collisions}
One of the initial ALICE goals in the pp run is the fine tuning of the detector
Monte Carlo simulation. Although a first successful commissioning was performed
with cosmic ray data, hadronic collisions offered the possibility to make a step forward in the detector 
understanding and Monte Carlo modelling.
As an example photon tagging allowed a measurement of the material budget radial distribution.
This gave the 
possibility to improve the detector knowledge and to obtain a much better 
detector simulation, where the $\gamma$ yield as a function of the distance from the IP 
nicely agree in the data and in the MC.
Proton runs are a rich source of physics, where ALICE exploits its peculiarity, 
making use of the low momentum cut.\\ 
{\bf - Multiplicity studies}\\
Multiplicity studies provide informations on the
energy density of the interaction and is one of the primary information needed to test 
Monte Carlo simulation. Moreover this is a basic variable,
so that a quick comparison between the four LHC experiment is made available. 
ALICE measured the charged particle multiplicity per 
pseudorapidity interval d$N$/d$\eta_{ch}$ at $\sqrt{s}$=0.9,~2.36 and 7 TeV, with the 
average charged particle multiplicity per unit of rapidity 
ranging from (3.81 $\pm$ 0.01 (stat) $\pm$ 0.07 (sys)) 
to (6.01 $\pm$  0.01 (stat) + 0.20,-0.12 (sys)).
While data agree nicely with the other LHC detectors, none of the investigated
models (Pythia, Phojet) and their tunes describe the average multiplicity and the
multiplicity distribution well.
At $\sqrt{s}$=0.9 and 2.36 TeV, the Pythia tunes Perugia-0 and D6T fail in 
reproducing the average multiplicity, while Phojet does not reproduce data
at $\sqrt{s}$= 7 TeV and is far away from describing the increase in multiplicity 
from $\sqrt{s}$=0.9 TeV to $\sqrt{s}$=2.36 TeV and from $\sqrt{s}$=2.36 TeV $\sqrt{s}$=7 TeV.
At 0.9 TeV, the high-multiplicity tail of the distributions is
best described by the Phojet model, while at 2.36 TeV,
Pythia tune ATLAS-CSC is the closest to the data.

The situation does not improve when considering the particle $p_{\rm t}$ prediction:
the transverse momentum
distribution at 900 GeV and the dependence of average $p_{\rm t}$ on $N_{ch}$ is not reproduced by the 
ATLAS-CSC Pythia tune [10]. At present we do not have an event generator/tune 
that can reproduce the
LHC data in a satisfactory way.\\ 
{\bf - Strange baryons}\\
The yields and $p_{\rm t}$ spectra of identified charged 
particle ($\pi$,~\rm{K},~\rm p) and neutral strange
particles ($\rm {K^{0}}$, $\Phi$, $\Lambda$, $\Xi$) have been measured at $\sqrt{s}$=0.9 and 7 TeV. 
While the $\phi$ is properly reproduced by Pythia (especially by the 
D6T tune), the $\rm K^{0}$ transverse
momentum spectrum is overestimated by the Pythia
tune ATLAS-CSC and Phojet below 0.75 GeV/$c$ but is
lower by a factor of $\simeq$2 in the $p_{\rm t}$ range 1-3 GeV/$c$. 
As far as strange baryons is concerned,
Phojet and Pythia tunes are well below the
data by a factor ranging from 3 to 10, depending on the baryon and on the particle $p_{\rm t}$.
Moreover data taken at $\sqrt{s}$=7 TeV shows the ratio $\Omega$/$\Xi$ is 
underestimated by Pythia of a factor up to 6.
From an experimental point of view it's worth noting the ratio of $\Lambda$/$\rm K^{0}$  
agrees very well with the STAR data taken at $\sqrt{s}$=200 GeV
and the ratio $\Xi$/$\Lambda$ is within the error.\\
{\bf - J/$\psi$ study }\\
J/$\psi$ study has been one of the most compelling evidence for quark gluon plasma
formation more than two decades ago. 
The study of this vector meson suppression at higher energy 
allows a big step in the understanding of the color field mechanisms at work in this new state of the
matter. Proton-proton runs offer the possibility to test the detector performance 
in J/$\psi$ detection and a reference data for Pb-Pb analysis.
ALICE can detect the  
 J/$\psi$$\rightarrow$
$\mu^{+}\mu^{-}$ channel 
 taking advantage of the forward
muon arm detector (2.5 < y < 4) and  
the  J/$\psi$$\rightarrow$
$e^{+}e^{-}$ channel by using the barrel detectors (|$\eta$| < 0.9).
ALICE measurement at central rapidity reaches $p_{\rm t}$ = 0 and is
therefore complementary to the CMS data, available at |y| < 1.2 for $p_{\rm t}$ > 6.5 GeV/$c$, and ATLAS,
which covers the region |y|<0.75 and $p_{\rm t}$> 7 GeV/$c$.
At $\sqrt{s}$=7 TeV the ALICE measured cross section is\cite{jpsi}:
\begin{equation}
\begin{split}
\sigma_{J/\psi}
(|\rm {y}|<0.9)=10.7 \pm 1.2 (sta.)\pm1.7 (sys.)+\\
+1.6 (\lambda= 1) -2.3(\lambda=-1) \mu b 
\end{split}
\end{equation}
\begin{equation}
\begin{split}
\sigma_{J/\psi}(2.5<\rm {y}<4)=6.31\pm0.25(sta.)\pm0.80(sys.)+\\
+ 0.95 ( \lambda=1)
-1.96 (\lambda=-1) \mu b,
\end{split}
\end{equation}
where $\lambda$ = 1
is for fully transverse and $\lambda$=-1 for longitudinal 
polarization.
The J/$\psi$ decaying into muons are compared to those 
detected by LHCb at 2.5 < y < 4, finding a good agreement. 
In the barrel region the CMS data (|y| < 2) and ATLAS (|y| < 0.9) can
be compared with those detected by ALICE only for $p_{\rm t}$ > 7 GeV. 
It is worth noting these results refer to inclusive production, therefore
the measured yield is a superposition of a direct component and of 
J/$\psi$
coming from the radiative decay of
higher-mass charmonium states.
\section {Results from Pb-Pb collisions}
Data collected during the 2010 gave a first
look at the hot and dense medium formed at $\sqrt{s_{\rm NN}}$=2.76 TeV when 
Pb-Pb ions collide.\\
{\bf - Energy density}\\
The energy density available in the Pb-Pb interactions 
is much larger with respect to the p-p one, resulting
in a very high number of particle produced. 
At $\sqrt{s_{\rm NN}}$=2.76 TeV and 
for central collisions, 
ALICE measured an average density of primary charged particles at
midrapidity <$N_{ch}$>=(1584$\pm$4(stat)$\pm$76(sys)). Normalizing 
per participant pair, we obtain
d$N_{ch}$/d$\eta$/(0.5$N_{part}$)=(8.3 $\pm$0.4(sys)),
about a factor 2 higher with respect to RHIC.
This is larger than most of
the predictions and about 50$\%$ more than expected from simple phenomenological
extrapolations from RHIC energy: the logarithmic law that described the multiplicity 
dependence with energy, does not hold anymore.
Following 
the Bjorken approach the average energy density has been derived.
The average amount of transverse energy produced per unit of pseudorapidity per participant pair
in central collisions is about 9 GeV, a factor $\simeq$3 larger than at RHIC (the
larger multiplicity at LHC being accompanied by an increase in the average transverse
momentum of the produced particles), corresponding to an energy density of about
15 GeV/$\rm fm^{3}$.
The centrality dependence of the charged particle multiplicity is rather mild, favouring
models incorporating some mechanism (such as parton saturation) moderating the
increase with centrality of the average multiplicity per participant pair.\\
{\bf - Nuclear modification factor  $R_{AA}$}\\
The partons generated by a ion-ion collision at high energy,
experience high energy loss collisions in the hot dense medium, 
showing a high opacity to their traveling inside. The depletion 
in the hadron yield is a powerful probe to investigate this effect. The nuclear modification factor
$R_{AA}$ is defined as the ratio
of the charged particle yield in Pb-Pb to that in pp, scaled by the
number of binary nucleon-nucleon collisions $N_{coll}$.
ALICE measured the nuclear modification factor
$R_{AA}$ of inclusive charged particle momentum distributions out to $p_{\rm t}$=20 GeV/$c$, where the
spectra are dominated by leading jet fragments. 
ALICE performed a first analysis\cite{RAA} where the prediction at $\sqrt{s}$=2.76 TeV was
extrapolated from the data collected at  $\sqrt{s}$=0.9 and 7 TeV. The analysis was improved
after data at  $\sqrt{s}$=2.76 TeV where taken. The two analysis agree quite well within the
systematical error and show the $R_{AA}$
ratio has a minimum at around 6 GeV, where the suppression is stronger than
at RHIC ( $\sqrt{s_{\rm NN}}$= 0.2 TeV), and then rises smoothly towards higher momentum. 
This latter feature
is not evident has not been seen in the published RHIC data. However, initial state effects
(shadowing/saturation), which presumably are very strong at LHC and which might
depend on both impact parameter and momentum transfer, can complicate a straight
forward interpretation of the data and the comparison between different beam energies.
The powerful ALICE PID allows the study of $R_{AA}$ for different hadrons separately.
\begin{figure}[h!]
\begin{center}
\includegraphics[width=5cm]{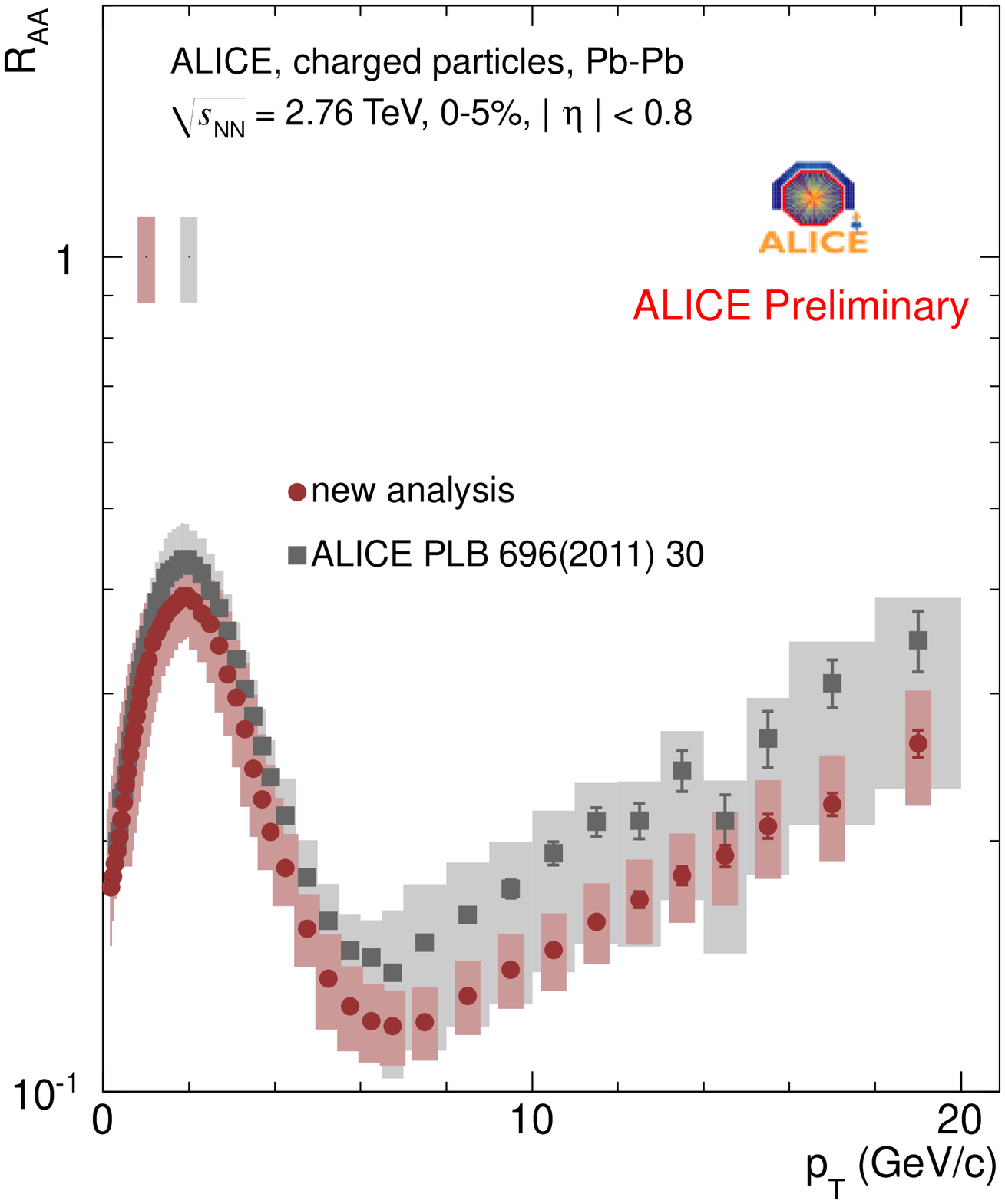}
\includegraphics[width=5cm]{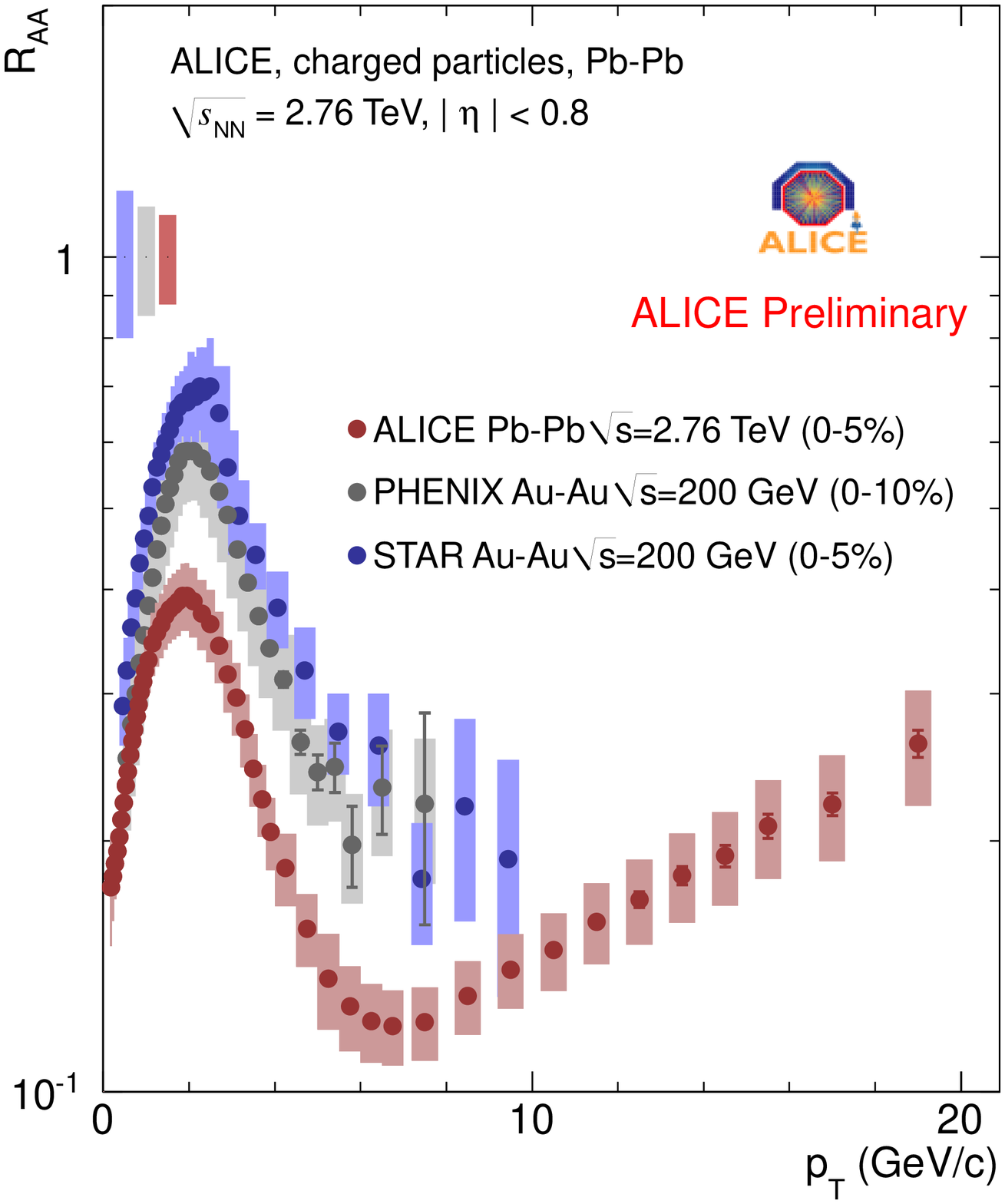}
\end{center}
\vspace{-0.5cm} \caption{ALICE $R_{AA}$ for the two different analysis(left) and compared to 
RHIC results(right)} \vspace{-0.5cm}
\label{fig:2}
\vskip 0.5 cm
\end{figure}
$R_{AA}$ looks
almost universal for $p_{\rm t}$>6 GeV/$c$; at low $p_{\rm t}$ the $\Lambda$ baryon don't show
any nuclear modification factor ($R_{AA}\simeq$1 ) while K behaves like all the other 
hadrons. It is worth noting $\rm D$ mesons are expected to show a smaller nuclear modification 
factor, since 
the main source of energy loss (gluon radiation) is depleted by the Casimir effect for heavy quark. 
This is found, although with a still high statistical error, in the data where a larger $R_{AA}$ 
is found for $\rm D^{0},D^{+}$ in the interval 4 GeV/$c$$\leq$$p_{\rm t}$$\leq$5 GeV/$c$.\\
{\bf - Elliptic flow}\\
The elliptic flow represents a powerful test to investigate the hydrodinamical properties 
of the quark gluon plasma. A perfect fluid shows a very small viscosity: this can
be studied by
looking at the efficiency in transferring the geometrical collision system anisotropy into 
momentum anisotropy. 
The distribution of the azimuthal angle, measured with respect to the reaction plane, is 
expanded 
into Fourier terms, where the second coefficient is the so called ``elliptic flow'', $v_2$.
The large elliptic flow observed at RHIC, is described reasonably well
by theoretical models based on relativistic hydrodynamics with a
QGP equation of state and a ratio of the shear viscosity to the 
entropy density
within a factor of $\simeq$5 by the supposed universal lower bound of 1/4$\pi$.
This indicates the QGP expands as a nearly perfect
fluid. The first ALICE results\cite{elliptic} shows 
the elliptic flow at  $\sqrt{s_{\rm NN}}$=2.76 TeV  is $\simeq$30$\%$ larger with respect to RHIC. 
Nevertheless the $v_2$ as a function of $p_{\rm t}$ is close to the RHIC measurement, showing 
the system hydrodynamic properties at RHIC and LHC are similar.
The increase of
the elliptic flow observed at LHC therefore comes from the increase of the average $p_{\rm t}$.
An important difference with respect to RHIC results is the elliptic flow study 
for different hadrons separately. While at RHIC the $v_{2}$/$n_q$, where $n_{q}$ is the number of hadron
valence quark, is similar for pion, kaons and protons, at LHC just pion and kaons $v_{2}$/$n_q$ is
compatible; protons have a lower $v_{2}$/$n_q$, showing the quark scaling does not hold 
for $p_{\rm t}$ < 0.5 GeV/$c$.\\ 
{\bf- J/$\psi$ suppression}\\
J/$\psi$ measurement is one of the key measurement for a high energy heavy ion experiment.
For $p_{\rm t}$>0 and 2.5<y<4, ALICE tags the J/$\psi$ through the  J/$\psi\rightarrow\mu^{+}\mu^{-}$ 
channel.
A  rather small J/$\psi$ suppression of about 0.5 was observed, practically independent of
centrality: this is a smaller suppression than that observed at RHIC.
An interesting result is the comparison with ATLAS, where data are taken only
at $p_{\rm t}$ > 6.5 GeV/$c$, shows a much stronger centrality dependence and suppression,
hinting for a $p_{\rm t}$ dependence of the J/$\psi$ suppression.
The measurement of the  J/$\psi$  in the   J/$\psi\rightarrow \rm {e^{+}e^{-}}$ channel is challenging
with the present statistics and large hadronic background. However the signal has been
extracted in the centrality class 0-40$\%$ and the central to peripheral 
(40-80$\%$) ratio ($R_{CP}$) has been evaluated. Within the large systematic uncertainties, 
the dielectron $R_{CP}$ is compatible with ATLAS and ALICE di-muon $R_{CP}$ measurements. 

\begin{figure}[ht!]
\begin{center}
\includegraphics[width=7cm]{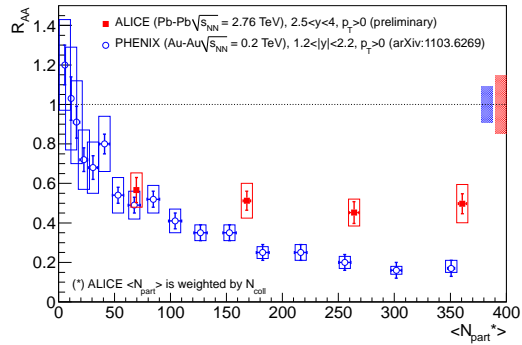}
\end{center}
\vspace{-0.5cm} 
\caption{ALICE J/$\psi$ measurement in the forward region compared with RHIC results.} 
\vspace{-0.5cm}
\label{fig:3}
\vskip  0.5 cm
\end{figure}

The above results hint at J/$\psi$ regeneration in hot matter at LHC energies, but
it is worth noting the J/$\psi$   production can be modified by the initial state effect
which could modify the medium: ALICE needs a p-Pb run as reference 
to disentangle the contributions from cold nuclear matter.\\

\end{document}